\documentclass{ws-procs9x6}
\usepackage{graphicx,epsfig,longtable,graphicx,subfigure,color,pifont}
\begin{document}
\title{THE DRIFT DIRECTIONAL DARK MATTER DETECTOR AND FIRST STUDIES OF THE 
HEAD-TAIL EFFECT}
\author{N.J.C. SPOONER and P. MAJEWSKI 
(for the DRIFT II Collaboration)}
\address{Department of Physics and Astronomy, University of Sheffield,
Sheffield, UK\\
$^*$E-mail: P.Majewski@sheffield.ac.uk, N.Spooner@sheffield.ac.uk\\}

\begin{abstract}
Measurement of the direction of the elastic nuclear recoil track and ionization charge distribution along it,
gives unique possibility for unambiguous detection of the dark matter WIMP particle. Within current radiation detection 
technologies only Time Projection Chambers filled 
with low pressure gas are capable of such measurement. Due to the character of the electronic and nuclear stopping 
powers of low energy nuclear recoils in the gas, an asymmetric ionization charge distribution along their tracks 
may be expected. Preliminary study of this effect, called Head-Tail, has been carried out here using the SRIM 
simulation program for Carbon and Sulfur in 40 Torr carbon disulfide, as relevant to the DRIFT detector. 
Investigations were focused on ion tracks projected onto the axis of the initial direction of motion 
in the energy range between 10 and 400 keV. Results indicate the likely existence of an asymmetry influenced by two competing 
effects: the nature of the stopping power and range straggling.  The former tends to result in the 
Tail being greater than the Head and the latter the reverse. It has been found that for projected tracks the mean 
position of the ionization charge flows from 'head' to 'tail' with the magnitude depending on the ion type and its energy.
\end{abstract}

\bodymatter
\section{Introduction} 

It has been demonstrated by the Dark Matter searching experiment DRIFT~\cite{drift1} using Time Projection Chambers (TPC) 
filled with low pressure CS${_2}$ that the direction of few mm long nuclear recoil tracks can be measured. However, due 
to the limited spatial resolution of the readout and ionization charge diffusion in the large detector volume, 
its distribution along the low energy nuclear recoil track is difficult to reconstruct. 

In this work preliminary results of simulations of the energy loss and ionization charge distribution along tracks of Carbon 
and Sulfur ions in 40 Torr CS${_2}$ ($\rho$=1.67$\cdot$10$^{-4}$ g/cm${^3}$) are presented. The so-called Head-Tail effect, 
describing the location of the majority of the ioinisation charge either at the beginning (Tail) or at the end (Head) 
of the track as a function of ion energy and diffusion has been studied. SRIM~\cite{srim} and TRIM programs were
used to calculate stopping powers for ions in gases, to investigate with Monte Carlo generators the energy loss and 
ionization charge distribution along the ion track on an event-by-event basis.

\section{SRIM}

\begin{figure}[ht]  
  \centering
  \psfig{file=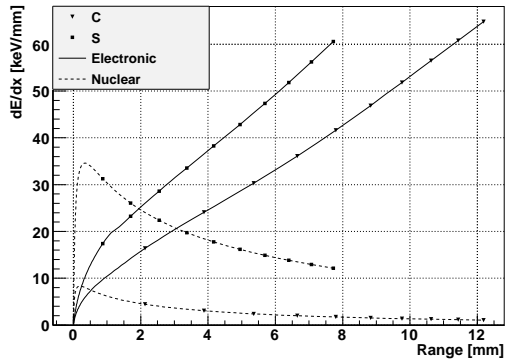,width=7.5cm,}
  \caption{Calculated with SRIM the ionization energy loss dE/dx as a function of ion range 
in the medium. Energy losses in electronic (solid line) and nuclear (dashed line) channels are 
shown for Sulfur(\ding{110}) and Carbon(\ding{116}) ions in 40 Torr CS${_2}$. The ion energy is 
marked up to 500 keV with an interval of 50 keV.}
  \label{f1}
\end{figure}

\begin{figure}[ht]
  \centering
  \psfig{file=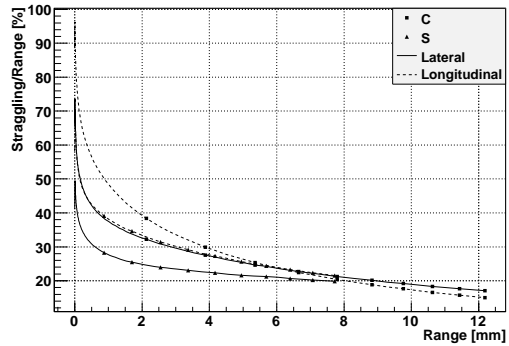,width=7.5cm,}
  \caption{Lateral (solid line) and longitudinal (dashed line) 
           straggling/range as a function 
           of range for Carbon (\ding{110}) and Sulfur (\ding{115})
           ions with energy is marked up to 500 keV with an interval 
           of 50 keV.}
  \label{f2}
\end{figure}

The Stopping and Range of Ions in Matter (SRIM) simulation program calculates as a function of energy: the 
electronic ($S_{e}$) and nuclear ($S_{n}$) energy loss, range and  longitudinal and lateral straggling of ions 
moving in matter. They have been calculated here for ion energies up to 500 keV. To understand the behaviour along the 
tracks they are plotted in Fig.~\ref{f1} and Fig.~\ref{f2} against ion range, the ion energies marked here on the 
plot at 50 keV intervals. 

\vspace{-0.9cm}

\vspace{-0.3cm}
As can be seen at higher energies the electronic energy loss is dominant and decreases with energy where at 
lower values the nuclear energy loss becomes greater than the electronic. For example for Sulfur ions the nuclear energy loss starts 
to dominate already at an energy of 100 keV. This means that energy loss along the ion track is not continuously decreasing.  This has an 
impact on the density of the ionization charge along the track and makes the head-tail effect energy dependent.

In the nuclear interactions secondary nuclear recoils are created. Although 3-4 times higher energy is needed to create 
an electron-ion pair by nuclear recoils than by electrons, 
they are a source of ionization. 

\vspace{-0.7cm} 

\vspace{-0.3cm}
Due to continuous collisions with gas atoms the direction of the moving ions become deviated from the original path. This 
causes fluctuations  in the ion's range, described by the lateral and longitudinal straggling and depending on ion energy 
(see Fig~\ref{f2}). One can see that ions slowing down experience increasing straggling relative to the drift 
length they have ahead, which rises sharply at the very end of the track. 

Tables calculated by SRIM are used in the Monte Carlo generator program called TRIM. In addition to on-line histogramming of 
various quantities, TRIM results are also recorded event by event in two different formats. The first is included in the 
file called {\it COLLISON.TXT} generated either in quick simulation, where for every ion collision its current energy, 
position in 3D and type of secondary recoils with their initial energy is recorded, or in full nuclear recoil 
cascade simulation, where information about the energies and positions of all created nuclear recoils is added. TRIM creates 
also an output file called {\it EXYZ} which, though with superior spatial resolution to the other outputs, delivers information only
about ion energy loss along the track. The latter format has been chosen here for the Head-Tail effect analysis.

\section{TRIM Simulation and Results Analysis}
\label{trim}
For each ion with a given energy 10$^4$ events were recorded and further analysed. The energies used were
for Sulfur: from 10 to 100 with a step of 10 keV then up to 250 with a step of 25 keV and up to 400 with a step of 50 keV 
and Carbon from 10 to 100 with a step of 10 keV then up to 200 keV with a step of 25 keV and up to 300 with a step of 
50 keV. Each event in the TRIM output {\it EXYZ} file contained on average 100 collision points with information about: current ion 
energy, the 3D coordinate of the collision point and energy loss, for which the sum equals the initial energy of the ion. 
For Carbon and Sulfur ions the number of ionization charges was calculated using W-values taken from~\cite{ifft}.
For each ion, a W-value was used at every collision according to its current energy. Note that the actual value of W 
at energy below 20 keV is not well known and values used here are based on an extrapolation of the function obtained 
from the polynomial fit to the existing data.

Based on this approximation the Head-Tail effect has been studied for three different configurations of ion interaction 
points: along the track, projected onto the line between the beginning and the end of the track and projected onto 
the axis of the initial ion direction. Because of the range straggling of ions with the same energy, the position
of each interaction projected point has been normalised to the projected range of the ion.

\vspace{-0.5cm}
\begin{figure}[ht] 
  \centering
  \psfig{file=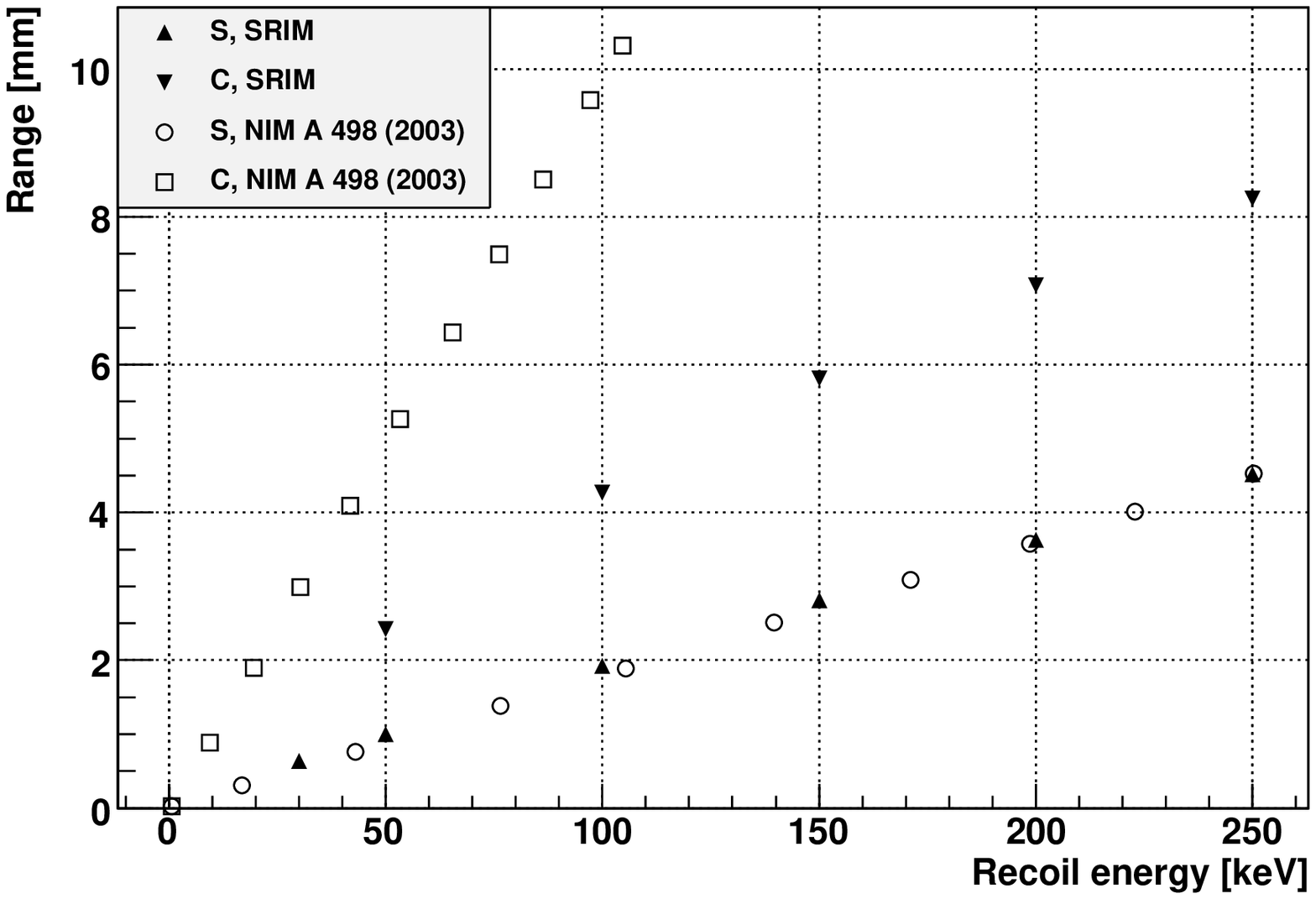,width=7.5cm,}
  \caption{Sulfur and Carbon ion range in CS$_2$ as a function of the energy, calculated with SRIM (\ding{115} and \ding{116}) 
   and taken from~\cite{ifft} (\ding{109} and \ding{111}).}
  \label{f5}
\end{figure}

\vspace{-0.5cm}
\subsection{Ion Ranges}

Using coordinates of the first and the last point of the ion track its projected range has been 
calculated. Fig.~\ref{f5} shows results for Sulfur and Carbon ions in CS${_2}$ in comparison with data 
from ~\cite{ifft}. There is very good agreement for Sulfur but a large discrepancy for Carbon ions for which results from this 
paper are more than twice smaller than in~\cite{ifft}.

\begin{figure}[ht] 
  \centering
  \psfig{file=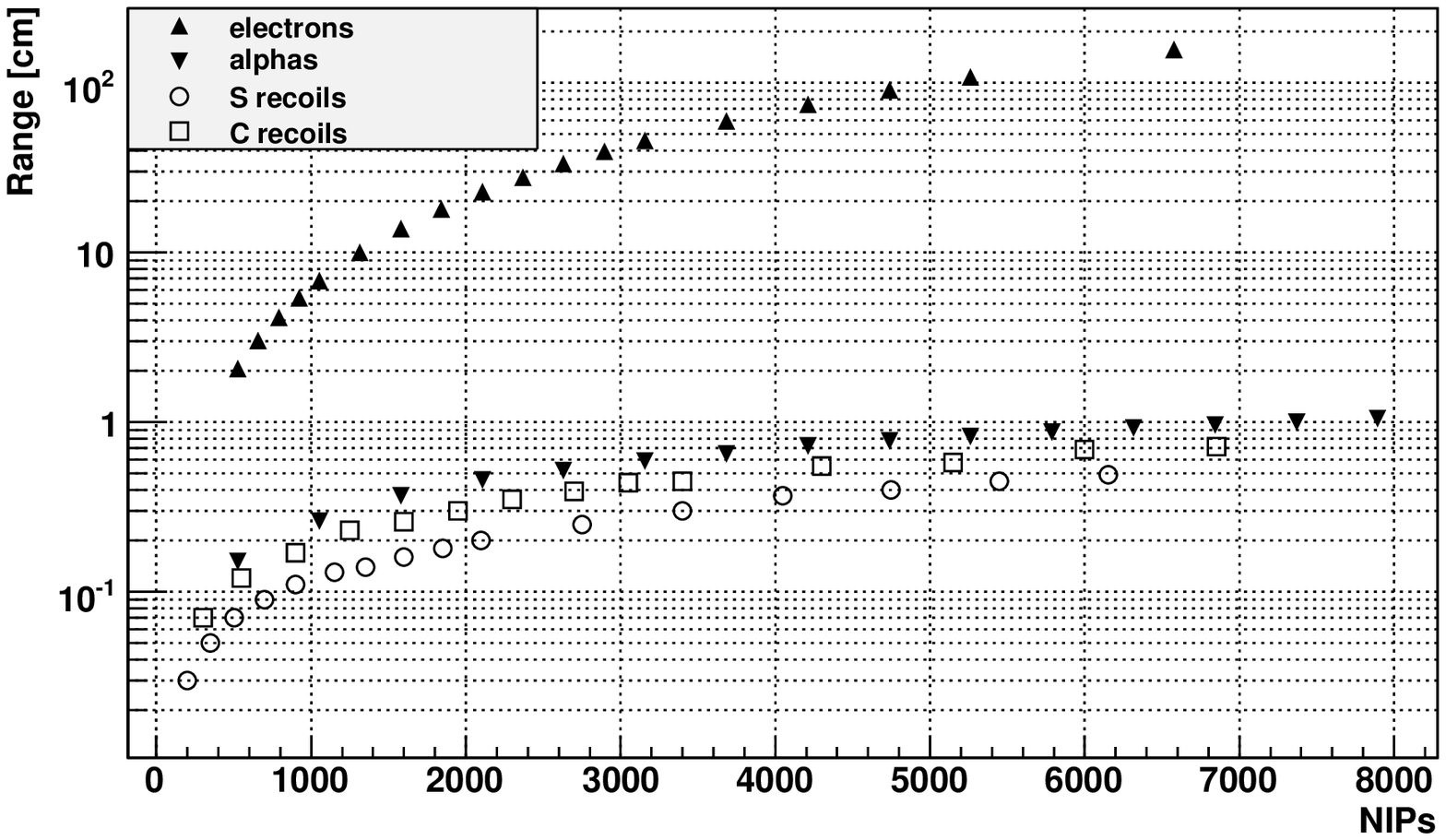,width=8.0cm,}
  \caption{Sulfur (\ding{109}), Carbon (\ding{111}), alpha (\ding{116}) and electron (\ding{115}) range in  CS$_2$ as a function of 
           ionization charge (NIPs) created. NIPs for Sulfur and Carbon were calculated using the W-values from ~\cite{ifft} while 
           for alpha particles and electrons a constant value of W=19~eV was used. To calculate the range of electrons the ESTAR program 
           was used for energies from 10 to 125 keV with steps 2.5, 5 and 10  up to 20, 60 and 100 keV, respectively. 
           Results for the alpha particle ranges are for energies from 10 to 150 with a step of 10 keV.}
  \label{f6}
\end{figure}

In the conversion of the energy loss into ionization charge, the number of electrons attached with no loss to CS${_2}$ molecules, 
creating CS${_2}^-$ and called Negative Ions Pairs (NIPs), was summed for each Carbon and Sulfur ion energy. Fig.~\ref{f6} shows, 
as a function of NIPs, the range of nuclear recoils together with the range of alpha particles 
and electrons calculated with TRIM and ESTAR~\cite{estar}, respectively. 

\vspace{-0.5cm}

\vspace{-0.9cm}
\begin{table}[ht]
\tbl{Ranges of 103 keV $^{206}$Pb recoils in pure and binary gases 
     at STP measured and simulated with TRIM. Values from measurement 
     are taken from~\cite{cano}. 1$\sigma$ of the simulated 
     range distributions is 10\% of the tabulated mean values 
     for all gases except Xe for which 
     1$\sigma$ is 15\%. Measurement uncertainty is $\pm$ 2$\mu$m.}
{\begin{tabular}{@{}ccccccc@{}}\toprule
Gas         & Ar & Xe & CH$_{4}$ & C$_{2}$H$_{4}$ & Air & N$_{2}$ \\
Measurement [$\mu$m] & 79 & 44 & 84       & 58             &83   & 80 \\
SRIM   [$\mu$m]     & 73 & 36 & 95       & 61             & 80  & 74  \\
\hline
\end{tabular}}
\label{tab1}
\end{table}

\vspace{-0.3cm}
The NIP values were calculated using W-value=19 eV. As one can see the range of electrons with the same number of NIPs as 
nuclear recoils is one order of magnitude greater than the range of recoils. This demonstrates that a TPC operating with 40 Torr CS${_2}$ has 
great separation power between electron and nuclear recoils. However, it shows also that the range of nuclear recoils and alpha particles 
are close to each other, with a difference not greater than 1-2 mm for tracks a few mm long, often also
elongated due to the charge thermal diffusion.

To check the consistency of the SRIM results with existing measurements, the range of 103 keV $^{206}$Pb recoils in several pure 
and binary gases at STP was calculated and compared with results from~\cite{cano}. The results are shown in Table ~\ref{tab1} 
indicating in most cases less than a 10\% difference. Similar differences between measurements and simulation were 
observed for 5-6 MeV alpha particles from Rn and Po decays in 40 Torr CS${_2}$~\cite{driftalpha}.

\subsection{Head-Tail}

\begin{figure}[ht!] 
  \centering
  \psfig{file=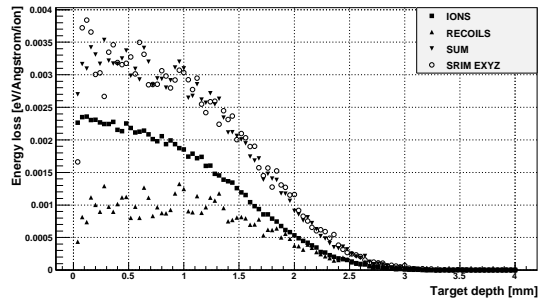,width=8.0cm,}
  \caption{Average ionization energy loss for 100 keV Sulfur ions (\ding{110}) and those created in the cascade secondary 
nuclear recoils (\ding{115}) as a function of target depth, obtained from TRIM histrogramming. 
Total energy loss (\ding{116}) is compared with scaled results from the {\it EXYZ} TRIM output file analysis (\ding{109}).}
  \label{f7}
\end{figure}

Along with the event-by-event information on ion energy loss, TRIM also delivers histograms accumulated on-line 
and showing distributions of average values of several quantities as a function of target depth. 
With default initial
TRIM settings the target depth corresponds to projection onto the axis of the initial direction of ion motion. 
One of the histograms shows distributions of the average ionization energy loss by ion and secondary recoils. 
An example of such a distribution, for 1000 Sulfur ions with energy of 100 keV, is shown in Fig.~\ref{f7}. 
One can 
see ionization energy loss by ion, secondary recoils and by the sum of them. The latter is compared with scaled 
results from the {\it EXYZ} TRIM output 
file analysis. Both show the same character of the ionization energy loss along the target depth indicating here 
that more energy is lost at the beginning of the track (Tail) than at the end (Head). However, this distribution is along 
the target depth regardless of the ion range which, due to straggling, varies from ion to ion. This is likely not the case in an 
actual experiment where recoils are detected on an event-by-event basis and charge distribution along the track is measured
with respect to its projected range. Thus for this work we study the energy loss and Head-Tail effect also using 
the {\it EXYZ} output file.

For Sulfur and Carbon ions the distributions of ionization charge were studied along: the track, projection onto a line 
between the first and last point of the track, and projection onto the axis of the initial ion motion. The latter were 
studied with and without diffusion. Diffused NIPs were redistributed in space according to the thermal diffusion 
formula: $\sigma=\sqrt{\frac{2k_{B}T}{e}}\sqrt{\frac{L}{E}}$ where: $k_B$ is Boltzmann constant, {\it T} ambient 
temperature, {\it e} electron charge, {\it E} drift electric field and {\it L} drift distance. After numerical 
simplifications a diffusion formula was used of form: $\sigma[mm]=0.72\sqrt{\frac{L[m]}{E[kV/cm]}}$ for three different drift 
lengths {\it L}: 0.1, 0.3 and 0.5 m and electric field in the DRIFT IIb detector {\it E}= 0.58 kV/cm.

\begin{figure}[ht] 
  \centering
  \epsfig{figure=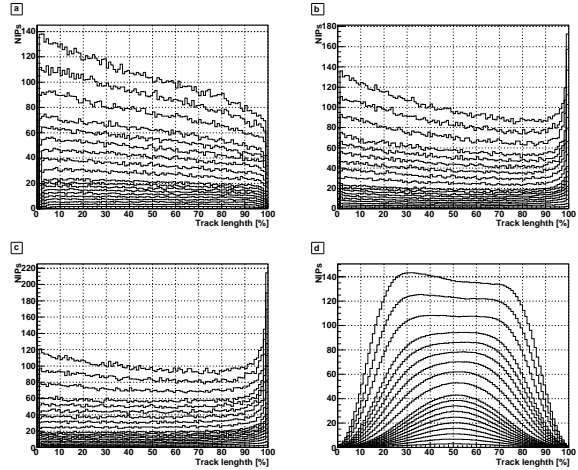,width=8.0cm,}
  \caption{Ionization charge distribution along Sulfur track (normalised to its range): (a) along the drift path, (b) projected onto the 
   line between the beginning and end of the track, (c) projected onto the direction of the initial motion, (d) randomly diffused with 
   maximum drift path of 50 cm and projected onto the direction of the initial motion. Displayed distributions are 
   for the following ion energies: from 10 to 100 with a step of 10 keV then up to 250 with a step of 25 keV and up to 400 with a step of
 50 keV.}
  \label{f10}
\end{figure}

\begin{figure}[ht]
  \centering
  \psfig{file=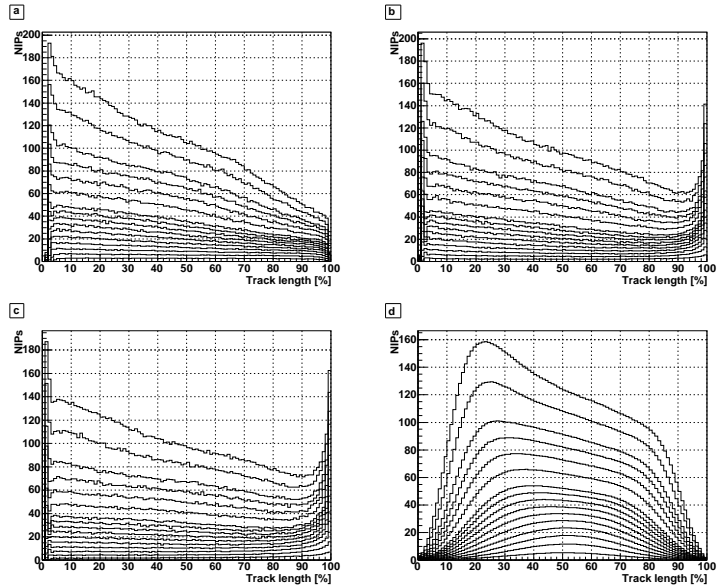,width=10.0cm,}
  \caption{The same as in Fig~\ref{f10} for Carbon tracks}
  \label{f11}
\end{figure}

Figs.[\ref{f10}] and [\ref{f11}] show NIP distributions along the Sulfur and Carbon tracks, respectively. For both, as expected, 
the ionization charge distributions along the line following the ion's motion (a) are not featured with a sharp rise at the 
end caused by the straggling. The negative slope of the distributions is very clear down to 40 keV and 20 keV for S and C, 
respectively from which at lower energies (where nuclear energy loss dominates) they become flat. 

Intermediate distributions of the charge projected onto a line between the track start and end point were studied. As one can see 
in (b) these still have negative slope but the sharp rise at the end is now clearly visible. The shape of distributions 
with projection onto the axis of the initial direction of motion is shown in (c). For Sulfur this is almost flat with a 
very sharp rise at the end, while for Carbon it remains with negative slope. An example of the impact of the diffusion caused 
by 0.5 m long drift, the maximum likely in DRIFT II, can be seen in Figs.~\ref{f10}d and~\ref{f11}d. 

\begin{figure}[ht]
  \centering
  \psfig{file= 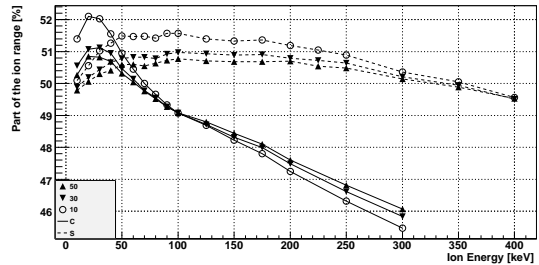,width=8.0cm,}
  \caption{Position of the mean of the charge distribution along Carbon (solid line) and Sulfur (dashed line) ion tracks showing the Head-Tail effect as a function of 
           energy and for different maximum drift distances: 10 (\ding{109}), 30(\ding{116}) and 50 (\ding{115}) cm. Points above 50 \% indicate that more ionization
           charge is created at the end (Head) while those below that level, at the beginning (Tail) of the track.}
\label{f12}
\end{figure}

For diffused distributions the Head-Tail effect was studied by calculation of the mean position of the distribution as a 
function of ion energy and drift length. Fig.~\ref{f12} shows the position of the mean along Carbon and Sulfur ion tracks 
with the head-tail effect dependent mainly on the ion energy. Points above 50\% are for ion tracks with greater 
`head` than `tail` and those below for ions with the reversed effect. 

\section{Conclusions}

In this work preliminary results of the Head-Tail effect of Carbon and Sulfur ions in low pressure CS$_{2}$ gas used in DRIFT detectors 
have been presented. Results were obtained using SRIM and TRIM simulation packages together with W-values available from existing 
data above ion energies of 20 keV. For lower ion energies W-values have been extrapolated using a polynomial fit to the existing 
data ranging from 20 to 300 keV. Results indicate the existence of the Head-Tail effect resulting respectively for Carbon and Sulfur 
greater Head for energies below 70 and 350 keV, above which Tail starts to dominate. As shown Head-Tail effect depends on the 
W-value which below 20 keV may be greater than values used in this work. This could impact significantly on the number of ionisation charge at the
Head of the track and could cause the Tail to become dominant at lower energy.

\end{document}